\documentclass[conference]{IEEEtran}
\IEEEoverridecommandlockouts
\usepackage{cite}
\usepackage{amsmath,amssymb,amsfonts}
\usepackage{algorithmic}
\usepackage{graphicx}
\usepackage{textcomp}
\usepackage{xcolor}
\def\BibTeX{{\rm B\kern-.05em{\sc i\kern-.025em b}\kern-.08em
    T\kern-.1667em\lower.7ex\hbox{E}\kern-.125emX}}

\usepackage[utf8]{inputenc}
\usepackage{amsthm}
\usepackage{amssymb}
\usepackage{verbatim}
\usepackage{bbm}
\usepackage{xcolor}
\usepackage{graphicx}
\usepackage{tikz}
\usetikzlibrary{positioning,chains,fit,shapes,calc}
\usepackage{blindtext}
\usepackage[margin=1in]{geometry}
\usepackage{amsmath}
\usepackage{amssymb}
\usepackage[english]{babel}
\usepackage{fancyvrb}
\usepackage{enumerate}
\usepackage{mathrsfs}
\usepackage{mathtools}

\newtheorem{theorem}{Theorem}[section]

\theoremstyle{definition}
\newtheorem{definition}{Protocol Rule}

\begin{document}

\title{Scalable Multi-Chain Coordination via the Hierarchical Longest Chain Rule\\
{\footnotesize}
\thanks{S. Vishwanath is an advisor to Dominant Strategies and incubates startups in the blockchain domain through ChainHub.}
}

\author{\IEEEauthorblockN{Yanni Georghiades}
\IEEEauthorblockA{\textit{Department of ECE} \\
\textit{University of Texas Austin}\\
Austin, TX \\
yanni.georghiades@utexas.edu}
\and
\IEEEauthorblockN{Karl Kreder}
\IEEEauthorblockA{Austin, TX \\
karl.kreder@gmail.com}
\and
\IEEEauthorblockN{Jonathan Downing}
\IEEEauthorblockA{\textit{Chief Architect} \\
\textit{Dominant Strategies}\\
Austin, TX \\
jonathan@dominantstrategies.io    }
\and
\IEEEauthorblockN{Alan Orwick}
\IEEEauthorblockA{\textit{CEO} \\
\textit{Dominant Strategies}\\
Austin, TX \\
alan@dominantstrategies.io}
\and
\IEEEauthorblockN{Sriram Vishwanath}
\IEEEauthorblockA{\textit{Department of ECE} \\
\textit{University of Texas, Austin}\\
Austin, TX \\
sriram@utexas.edu}
}

\maketitle

\begin{abstract}
This paper introduces BlockReduce, a Proof-of-Work based blockchain system which achieves high transaction throughput by means of a hierarchy of merged mined blockchains, each operating in parallel on a partition of the overall application state. Most notably, the full PoW available within the network is applied to all blockchains in BlockReduce, and cross-blockchain state transitions are enabled seamlessly within the core protocol. This paper shows that, given a hierarchy of blockchains and its associated security model, BlockReduce scales superlinearly in transaction throughput with the number of blockchains operated by the protocol.
\end{abstract}


\section{Introduction}
\label{sec:intro}

Blockchains are popular as a means to enable trustless, decentralized, peer-to-peer value transfer. 
Among the approaches to achieving distributed consensus in cryptocurrencies,  Proof-of-Work (PoW) is the oldest, most established, and arguably most well-understood.  
However, PoW-based cryptocurrencies are currently limited in terms of transaction throughput in comparison with traditional payment mechanisms such as credit cards.  
This has resulted in  increased transaction costs and a greater shift towards alternate scaling mechanisms.
In particular, Proof-of-Stake (PoS), other Proof-of-X consensus protocols, and a proliferation of Layer 2 protocols have been proposed and implemented in order to enable lower transaction fees. 
All of these approaches have different trust models in comparison with Proof-of-Work and, as a result, come with their own associated challenges and weaknesses.

The two most notable cryptocurrencies, Bitcoin and Ethereum, are both PoW-based\footnote{at the time of writing, Ethereum has not yet attempted the transition to Proof-of-Stake} and have a maximum throughput of under 20 transactions per second \cite{georgiadis2019many,chauhan2018blockchain}, whereas Visa alone can execute more than 2,000 transactions per second on their credit card network \cite{chauhan2018blockchain}. 
Indeed, it is now presumed (without proof) by a majority of people that PoW cryptocurrencies simply cannot meet the throughput requirements of a global currency. 

In this paper, we introduce BlockReduce, a PoW cryptocurrency that achieves high transaction throughput (as a Layer 1 protocol). 
We describe BlockReduce by first identifying the primary factors that cause low transaction throughput (and therefore, large fees) in PoW blockchains. 
We then address and ameliorate each factor, resulting in a truly scalable solution.

The primary tools underlying our solution for scalability are as follows:

\textbf{Latency-dependent Clustering of Network Nodes}: As noted across the literature \cite{wan2019evaluating}, network latency is one of the biggest factors in the scalability of blockchain systems. In our work, we translate this understanding of network latency into a suitable hierarchical clustering, where nodes self-partition into a hierarchy of sub-networks with which they share low-latency connections.
Each sub-network operates its own blockchain to validate and update a partition of the overall application state. 

\textbf{Transaction-dependent Security}:  Currently, a vast majority of PoW cryptocurrencies afford the same level of security---where, for our purposes, security refers to the amount of work an adversary would need to perform in order to succeed in a double-spend attack---for all transactions, regardless of the economic value of the transaction. 
However, in most human-commerce interactions, low-value transactions are not secured to the same level as high-value transactions. 
For example, credit card transactions of low value often do not require signatures, while higher value transactions go through a more stringent signature verification process. 
Ultimately, even in the blockchain domain, we believe that security should be transaction-value dependent, with high-value blocks (and associated transactions) afforded greater security guarantees. 
Thus, the amount of work applied to all transactions for the sake of settlement need not be the same in the short-term, although in order to prevent transaction conflicts from proliferating, eventually all transactions in BlockReduce must be validated and secured by the maximum amount of work available to the system.

\textbf{Merged Mining}: BlockReduce is composed of a tree of blockchains operated in parallel. 
Rather than performing PoW computations on a single block header, miners simultaneously mine a blockchain at each level of the tree using the same PoW computations, and one PoW solution might correspond to a block in multiple blockchains. 
This has two effects in BlockReduce.
The first is that all miners always mine on the root of the tree, meaning the root blockchain receives all of the mining power of the network.
The second is that blocks are found periodically which are shared between blockchains at different levels of the tree (these blocks are called \textit{coincident blocks}), which allows work to be shared across blockchains and also enables cross-blockchain state-transitions.





\subsection{Proof-of-Work Blockchains}
The first published instance of Proof-of-Work being applied to blockchains is Nakamoto's famous Bitcoin protocol \cite{Nakamoto2008}, where Nakamoto combined PoW with a block selection rule called the Longest Chain Rule to achieve Nakamoto consensus on transactions. 
BlockReduce uses PoW to reach a similar form of consensus on each blockchain in the tree. 
In order to accommodate for this hierarchical structure, we define a variant of the Longest Chain Rule which we refer to as the Hierarchical Longest Chain Rule.
We describe these aspects of the BlockReduce protocol in more detail in Sections \ref{sec:pow} and \ref{sec:hlc}. 

\subsection{Proof-of-Work Efficiency}
A significant limiter of transaction throughput in public blockchains is the amount of time it takes for data to propagate within the peer-to-peer network after blocks are mined. 
In order to provide intuition about this phenomenon, we define the \textit{PoW efficiency} of a blockchain as the fraction of PoW computations which contribute to the canonical chain (i.e., the chain of blocks which are committed to the transaction ledger/state machine).
In the ideal setting in which all nodes follow the protocol and there are no network delays on block propagation, the PoW efficiency, $\mathcal{E}$, is 1. 
This is because when there are no propagation delays, all miners instantaneously adopt each new block into their canonical chain. 

To model a more realistic setting, we define $\lambda$ as the total rate of block generation by the network and the network delay $\Delta$ as the time between when a block is found and when it is received by all nodes in the network.
Under this model, \cite{Pass2016} computes the effective block generation rate to be $\frac{\lambda}{1 + \lambda \Delta}$, resulting in a PoW efficiency of $\mathcal{E} = \frac{1}{1 + \lambda \Delta}$.

Furthermore, if a fraction $\beta$ of nodes are adversarial and are mining a private branch of the blockchain as part of an attack, then \cite{Pass2016} computes the PoW efficiency of honest nodes to be $\mathcal{E} = \frac{(1 - \beta)}{1 + (1 - \beta) \lambda \Delta}$, which varies inversely with $\Delta$.
On the other hand, assuming the adversary experiences negligible network delay with itself while mining a private chain (i.e., $\Delta \approx 0$), then their PoW efficiency is approximately 1.
Herein lies the problem. If $\Delta$ is fixed, then an adversary's relative advantage over honest nodes increases as the block generation rate increases. 
Systems such as Bitcoin suppress $\lambda$ in order to reduce the impact of $\Delta$ on the PoW efficiency of honest nodes. 
This phenomenon is discussed in various works \cite{bagaria2019prism, dembo2020everything}, and several solutions have been presented which assume $\Delta$ to be fixed. 
In BlockReduce, we partition the network into sub-networks so that the $\Delta$ experienced by each sub-network is smaller than that of the overall network, thereby allowing each sub-network to operate a blockchain with a higher block generation rate.
We can achieve this while maintaining an equal PoW efficiency across all blockchains, which we believe is critical to a truly scalable blockchain.
Moreover, we utilize the Hierarchical Longest Chain Rule described in Section \ref{sec:hlc} to guarantee that each blockchain receives the maximum amount of work available to the network, thereby achieving high transaction throughput without sacrificing security or PoW efficiency.

\subsection{Our Contributions}
We present BlockReduce, a PoW blockchain system which enables high transaction throughput by utilizing a hierarchy of merged mined blockchains operating in parallel on non-overlapping partitions of the application state. 
To the best of our knowledge, BlockReduce is the first protocol which promises superlinear scaling with the number of parallel blockchains it operates while still securing each blockchain with the maximum amount of work available to the network.
We introduce the Hierarchical Longest Chain Rule as a block-selection mechanism which allows each blockchain in the hierarchy to inherit the work of its parent blockchain and also enables native, cross-blockchain state transitions between any state partitions in the hierarchy. 
Finally, we analyze the performance of the protocol to demonstrate the superlinear scalability of BlockReduce. 

\section{Related Work}
\label{sec:related}


There have been a number of proposals for scaling transaction throughput in blockchains. 
We provide a brief summary of several types of approaches. 

\subsection{Parallel PoW Blockchains}
In a manner conceptually similar to BlockReduce, many protocols aim to achieve high transaction throughput by operating several blockchains in parallel.
The PoW version of Parallel Chains \cite{Fitzi2018} involves mining a metablock containing candidate blocks for a number of parallel chains which operate non-overlapping state partitions. 
Notably, Parallel Chains does not support cross-blockchain transactions and is therefore of more limited application than BlockReduce. 
Chainweb \cite{martino2018chainweb} is another protocol operating many parallel chains, where each block header references the headers of other chains in order to braid the chains together. 
Chainweb allows cross-blockchain state transitions and also features a mechanism by which chains inherit work from one another, but it can achieve only a linear increase in transaction throughput with the number of parallel chains whereas BlockReduce achieves superlinear scaling. 

\subsection{Proof-of-Stake Protocols}
Many Proof-of-Stake protocols have been proposed and implemented, such as Ouroboros Praos \cite{kiayias2017ouroboros} and Ethereum's planned move to Proof-of-Stake, as a means to enable high transaction throughput and low settlement times.
However, Proof-of-Stake protocols currently do not afford the same security guarantees as PoW and also often suffer from shortcomings such as the ``nothing at stake" problem or predictability on the next eligible validator \cite{brown2019formal}.

\subsection{Layer 2 Protocols}
There are multiple Layer 2 protocols (i.e., protocols which operate independently and only periodically interact with the blockchain) which have been developed in order to facilitate high transaction throughput. 
These include Starkware, Polygon, and Lightning among many others \cite{sguanci2021layer}. 
Although some such architectures can achieve high transaction throughput, Layer 2 solutions inevitably require alternate trust models from the core blockchain protocol which may not be suited for all use cases. 
BlockReduce scales as a Layer 1 protocol and does not require those additional assumptions to achieve high transaction throughput.




\section{Model}
\label{sec:model}
In this section, we describe the network model under which we analyze the BlockReduce protocol.
We adopt a simple {\em overlay}-network model to understand the interactions between nodes: that of a $d$-regular graph on $N$ network nodes.
This model arises from the standard set forth by Bitcoin, where the protocol executed by each node attempts by default to maintain a set of 8 peers.
We assume a synchronous, round-based model for block propagation. 
According to \cite{Fountoulakis2013}, with probability $1 - o(1)$, the number of synchronous rounds required for data to broadcast via gossip in a random $d$-regular network of $N$ nodes is 

\begin{equation}
    (1 + o(1))\big(\frac{1}{\ln (2(1 - \frac{1}{d}))} - \frac{1}{d \ln (1 - \frac{1}{d})} \big) \ln N.
\end{equation}

We use this result to characterize the overall network propagation delay $\Delta$ (i.e., the amount of time it takes for a single block to be propagated across the network) in terms of the average single-link propagation delay, $\delta$, as follows. 

\begin{equation}
\label{eq:delta}
    \Delta = \delta (1 + o(1))\big(\frac{1}{\ln (2(1 - \frac{1}{d}))} - \frac{1}{d \ln (1 - \frac{1}{d})} \big) \ln N.
\end{equation}


Adopting the shortened notation of \cite{Fountoulakis2013}, this gives us the following bound: 

\begin{equation}
\label{eq:delta_approx}
    \Delta < \delta C_d \ln N.
\end{equation}

We define $\Delta$ in this way to show that $\Delta$ decreases as $\delta$ or $N$ decrease.
In Section \ref{sec:hierarchy} we use this understanding to show that $\Delta$ is reduced in higher order blockchains within the BlockReduce protocol, thereby allowing for higher block generation rates while still maintaining high PoW efficiency within those blockchains.

\section{A Hierarchy of Blockchains}
\label{sec:hierarchy}
In this section, we describe the BlockReduce protocol and show that all nodes operating a particular blockchain are able to agree upon its state. 

\subsection{Notation}
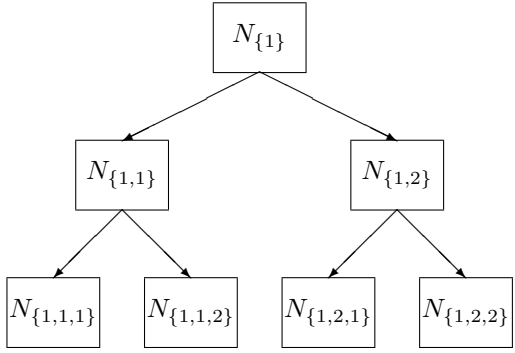
\begin{figure}[h]
\setlength{\unitlength}{0.12in} 
\centering 
\begin{picture}(24,16) 
\put(9,12){\framebox(4,3){$N_{\{1\}}$}}
\put(3,6){\framebox(4,3){$N_{\{1,1\}}$}}
\put(15,6){\framebox(4,3){$N_{\{1,2\}}$}}
\put(0,0){\framebox(4,3){$N_{\{1,1,1\}}$}}
\put(6,0){\framebox(4,3){$N_{\{1,1,2\}}$}}
\put(12,0){\framebox(4,3){$N_{\{1,2,1\}}$}}
\put(18,0){\framebox(4,3){$N_{\{1,2,2\}}$}}

\put(11,12){\vector(-2,-1){6}}\put(11,12){\vector(2,-1){6}}
\put(5,6){\vector(-1,-1){3}}\put(5,6){\vector(1,-1){3}}
\put(17,6){\vector(-1,-1){3}}\put(17,6){\vector(1,-1){3}}
\end{picture}
\caption{Hierarchy tree with 3 orders, where each box is a sub-network which operates a distinct blockchain.}
\label{fig:tree}
\end{figure}

\begin{figure}[htp]
    \centering
    \includegraphics[width=8cm]{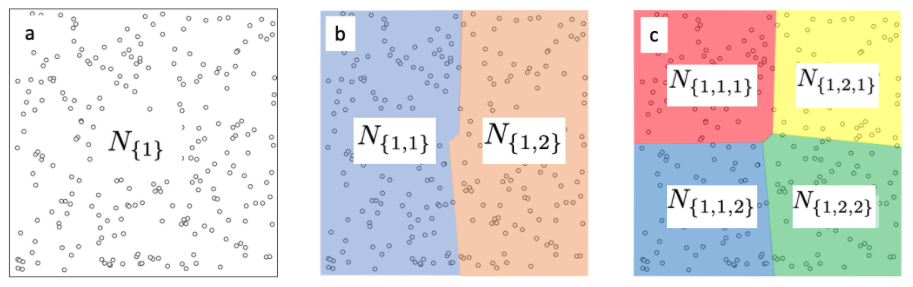}
    \caption{Illustration of topological network segregation: a) full network b) two order 2 sub-networks c) four order 3 sub-networks.}
    \label{fig:galaxy}
\end{figure}

The BlockReduce protocol requires that nodes self-partition into a hierarchy of sub-networks, where the hierarchy is a tree consisting of $R$ levels called \textit{orders}. 
Each sub-network is denoted $\mathcal{N}_{\vec{v}}$, where $\vec{v}$ is the unique path from the root to the specified node in the hierarchy tree. 
The root of the hierarchy tree is of order $1$, and the leaves are of order $R$. 
An example of a BlockReduce hierarchy structure and its associated sub-network partitioning is provided in Figures \ref{fig:tree} and \ref{fig:galaxy}. 
In this example, the hierarchy has three orders, and networks of order 1 and 2 each have two child sub-networks.

Each sub-network $\mathcal{N}_{\vec{v}}$ operates a blockchain $\mathcal{B}_{\vec{v}}$ to achieve consensus on a partition of the state $\mathcal{S}_{\vec{v}}$. 
Blockchains of order $r$ have block arrivals at a rate $\lambda_r$, network delay $\Delta_r$, and an average single-link propagation delay $\delta_r$.

We adopt a functional notation when discussing relationships between sub-networks, blockchains, and state partitions. 
We denote $\mathsf{parent}(\mathcal{N}_{\vec{v}})$ to be the parent sub-network of $\mathcal{N}_{\vec{v}}$ in the hierarchy tree and $\mathsf{order}(\mathcal{N}_{\vec{v}}) = |\vec{v}|$ to be the order of $\mathcal{N}_{\vec{v}}$.
We often overload this functional notation with inputs $\mathcal{B}_{\vec{v}}$, $\mathcal{S}_{\vec{v}}$, or simply $\vec{v}$, and these functions are evaluated in the same way.


\subsection{Merged Mining}
\label{sec:pow}
Mining in BlockReduce is similar to mining in a standard PoW blockchain except that, in BlockReduce, multiple blocks are mined simultaneously using a technique called \textit{merged mining} \cite{xu2017taxonomy}, and the Longest Chain Rule (LCR) introduced by Bitcoin \cite{Nakamoto2008} is modified to accommodate the hierarchical structure of BlockReduce.


Each BlockReduce miner selects an order $R$ blockchain (a leaf node in the tree) and simultaneously mines each blockchain along the path from root to leaf. 
For example, the miner might select the leaf $\mathcal{B}_{\{1, 1, 1\}}$, in which case they would simultaneously mine $\mathcal{B}_{\{1\}}$, $\mathcal{B}_{\{1, 1\}}$, and $\mathcal{B}_{\{1, 1, 1\}}$.
To accomplish this, miners construct a block for each blockchain they are mining, concatenate the block headers together, and perform PoW computations on the combined block header; as a result, those blocks will all share the same block hash.

\begin{figure}[h]
\setlength{\unitlength}{0.11in} 
\centering 
\begin{picture}(30,20) 

\put(0,2){\framebox(4,3){{\footnotesize 0x0FFF}}}
\put(6,2){\framebox(4,3){{\footnotesize 0x00FF}}}
\put(12,2){\framebox(4,3){{\footnotesize 0x0FFA}}}
\put(18,2){\framebox(4,3){{\footnotesize 0x000F}}}
\put(25,3){$\mathcal{B}_{\{1,1,1\}}$}

\put(6,3.5){\vector(-2,0){2}}
\put(12,3.5){\vector(-2,0){2}}
\put(18,3.5){\vector(-2,0){2}}

\put(6,7){\framebox(4,3){{\footnotesize 0x00FF}}}
\put(18,7){\framebox(4,3){{\footnotesize 0x000F}}}
\put(18,8.5){\vector(-2,0){8}}
\put(25,8){$\mathcal{B}_{\{1,1\}}$}

\put(18,12){\framebox(4,3){{\footnotesize 0x000F}}}
\put(25,13){$\mathcal{B}_{\{1\}}$}

\put(8,6){\oval(6,10)}
\put(20,8.5){\oval(6,15)}

\put(1.5,0){$B_1$}
\put(7.5,0){$B_2$}
\put(13.5,0){$B_3$}
\put(19.5,0){$B_4$}

\put(5.2, 13){\textit{Coincident}}
\put(6.5, 11.5){\textit{block}}

\put(17.2, 18){\textit{Coincident}}
\put(18.5, 16.5){\textit{block}}
\end{picture}
\caption{Example block visualization for a system with 3 orders. Each box represents a block, and the value inside the box is the hexadecimal representation of the block hash. Blocks in the bottom row make up blockchain $\mathcal{B}_{\{1,1,1\}}$ and so forth. Block $B_2$ meets the difficulty requirement of $\mathcal{B}_{\{1,1\}}$ and is therefore a coincident block shared by orders 2 and 3, $B_4$ is also a coincident block but is shared by all 3 orders, and blocks $B_1$ and $B_3$ only meet the difficulty requirement of $\mathcal{B}_{\{1,1,1\}}$.}
\label{fig:coincident}
\end{figure}
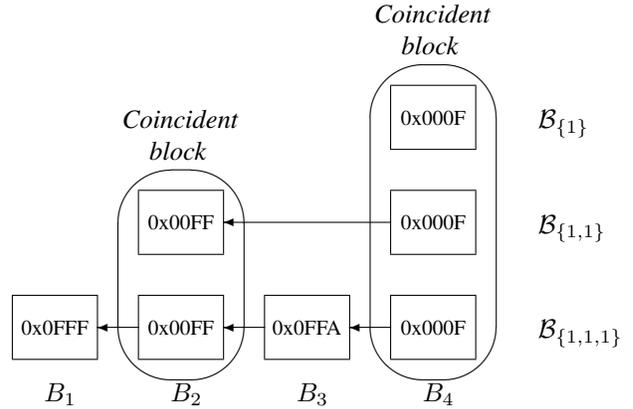

Blockchains closer to the root of the hierarchy have increasing and overlapping PoW difficulties so thaat a block which meets the difficulty requirement of a blockchain of order $r$ also meets the difficulty requirement of each blockchain of order greater (i.e., further from the root) than $r$. 
For example, Figure \ref{fig:coincident} depicts a sequence of blocks in a system with 3 orders.
In this example, the difficulty requirement for an order 1 block is that the block hash has 12 leading 0's (in the binary expansion), whereas an order 2 block requires only 8 leading 0's, and an order 3 block requires 4 leading 0's. 
In this case, an order 1 block with 12 leading 0's also meets the difficulty requirement of each other order.
A block which is shared by multiple orders in this way is called a \textit{coincident block} because it serves to coincide these blockchains under a shared block reference. 
In Section \ref{sec:hlc}, we show that under the Hierarchical Longest Chain Rule, coincident blocks impose a partial ordering on blockchains of different orders which enables cross-blockchain state transitions.
 
\subsection{Partitioning the Ledger State}
\label{sec:state}
We adopt a generic state model in which each transaction constitutes an update to the application state.
This generality allows BlockReduce to support the Unspent Transaction Output (UTXO) model, in which the application state is simply the set of all UTXOs, as well as more sophisticated smart contract models, where the application state is the smart contract state.
State is partitioned between all blockchains to prevent duplication, and for the purpose of cross-blockchain state transitions, each transaction must specify an origin blockchain $\vec{v_o}$ and a destination blockchain $\vec{v_d}$.
If the origin and destination of a transaction are the same, the state transition occurs in the same way as a traditional blockchain implementation. 
If the origin of a transaction is different from its destination, for example if an asset is being tranferred from one blockchain to another, then the state update involves removing the asset from the origin state and adding it to the destination state. 
This leads to the following protocol rule. 




\begin{definition}
In order for a transaction with origin $\vec{v_o}$ to be valid, the state that it modifies must be valid with respect to $\mathcal{S}_{\vec{v_o}}$.
\end{definition}

For example, if a user is attempting to move an asset from $\vec{v_o}$ to $\vec{v_d}$, then the transaction must have origin $\vec{v_o}$ and the user must demonstrate ownership of the asset in $\mathcal{S}_{\vec{v_o}}$.

\subsection{The Hierarchical Longest Chain Rule}
\label{sec:hlc}
BlockReduce utilizes a novel consensus rule to select the \textit{canonical chain}---i.e., the chain of blocks referencing state updates to be applied---for each blockchain. 
In Bitcoin and other more traditional systems, the Longest Chain Rule (LCR) stipulates that the canonical chain is the sequence of valid blocks with the most work (often referred to as the longest chain or, more accurately, the heaviest chain) \cite{Nakamoto2008}. 
BlockReduce follows a similar rule but must also account for the existence of coincident blocks within the hierarchy.

Before defining the \textit{Hierarchical Longest Chain Rule} (HLCR), we first define what conditions a block must meet to be considered valid. 
While specific requirements may vary between implementations, such as varying block size, transaction and/or smart contract structure, or block header composition, in general we can define a valid block as follows. 

\begin{definition}[Valid Block]
A block is considered valid if it meets all protocol rules and all of its predecessors (of any order) are also valid. 
\end{definition}

In other words, a valid block must conform to the rules of the blockchain and must reference no prior blocks which deviate from those rules. 
Importantly, because a coincident block has a predecessor at multiple orders, if \textit{any} of its predecessors are invalid then the block is also invalid. 
This guarantees that a coincident block is either valid in all blockchains for which it meets the difficulty requirement or none of them. 
Next, we define the HLCR which miners use to determine the canonical chain in BlockReduce.

\begin{definition}[The Hierarchical Longest Chain Rule]
\label{rule:hlc}
The canonical chain of the root blockchain $\mathcal{B}_{\{1\}}$ is the heaviest sequence of valid blocks in $\mathcal{B}_{\{1\}}$.
The canonical chain of any blockchain  $\mathcal{B}_{\vec{v}}$ of order greater than 1 is the heaviest sequence of blocks which contains \textit{all} coincident blocks between $\mathcal{B}_{\vec{v}}$ and $\mathsf{parent}(\mathcal{B}_{\vec{v}})$ which are present in the canonical chain of $\mathsf{parent}(\mathcal{B}_{\vec{v}})$, and \textit{no} coincident blocks between $\mathcal{B}_{\vec{v}}$ and $\mathsf{parent}(\mathcal{B}_{\vec{v}})$ which \textit{are not} present in the canonical chain of $\mathsf{parent}(\mathcal{B}_{\vec{v}})$. 
\end{definition}

In other words, the canonical chain for the root blockchain is selected via the standard LCR. 
For each other blockchain $\mathcal{B}_{\vec{v}}$ of order greater than $1$, the canonical chain must include all coincident blocks that are shared between $\mathcal{B}_{\vec{v}}$ and $\mathsf{parent}(\mathcal{B}_{\vec{v}})$ that are present in the canonical chain of $\mathsf{parent}(\mathcal{B}_{\vec{v}})$.
However, if there is some coincident block between $\mathcal{B}_{\vec{v}}$ and $\mathsf{parent}(\mathcal{B}_{\vec{v}})$ that is not in the canonical chain of $\mathsf{parent}(\mathcal{B}_{\vec{v}})$, then it cannot be in the canonical chain of $\mathcal{B}_{\vec{v}}$.

If an incoming block causes the canonical chain to change, then the state updates dictated by blocks which are no longer part of the canonical chain must be reverted and the new state updates applied.
This is why the canonical chain must contain all coincident blocks that are shared with the parent blockchain, as otherwise a cross-blockchain state transition which has been applied at both origin and destination could later be reverted at the origin but not the destination, thus causing an inconsistency in the overall network state.

\subsection{Inter-Blockchain Ordering via Coincident Blocks}
\label{sec:ordering}


Within a single blockchain, all blocks in the canonical chain are totally ordered according to their distance from the genesis block. 
Between blockchains of different orders, blocks are partially ordered due to the coincident blocks which arise from merged mining.

Intuitively, a coincident block serves as a shared point in ``time" between blockchains, allowing nodes to agree upon which blocks came ``before" the coincident block and which blocks came ``after."
For example, in Figure \ref{fig:coincident}, nodes in $\mathcal{N}_{\{1,1\}}$ which have received block $B_4$ can all agree that block $B_3$ came before $B_4$ even if they are not mining $\mathcal{B}_{\{1,1,1\}}$.


This property of coincident blocks allows all nodes in both $\mathcal{N}_{\vec{v}}$ and $\mathsf{parent}(\mathcal{N}_{\vec{v}})$ to agree on the existence and ordering of all blocks prior to the coincident block in either blockchain. 
In Section \ref{sec:state}, we show that this enables cross-blockchain state transitions to occur, as nodes operating the destination blockchain can agree on precisely if and when the state transition should be applied to the destination state.

\subsection{Inherited Work via Coincident Blocks}
\label{sec:shared_work}
We argue that, for the sake of short term settlement, blocks containing transactions of low economic value may be secured with a fraction of the maximum work available to the network, as the potential economic loss from a successful attack is small. 
However, in the longer term, this security level is not sufficient.
If a block containing a cross-blockchain state transition were to be removed from the canonical chain of $\mathcal{B}_{\vec{v_o}}$, but the state transition for that transaction had already been applied to $\mathcal{S}_{\vec{v_d}}$, then an inconsistency in the overall application state might arise.

The HLCR prevents this type of inconsistency by requiring that the canonical chain of a child blockchain must contain any coincident blocks that are shared with its parent blockchain, regardless of the number of blocks in any competing fork. 
In other words, a coincident block is removed from the canonical chain of $\mathcal{N}_{\vec{v}}$ if and only if it is first removed from the canonical chain of $\mathsf{parent}(\mathcal{N}_{\vec{v}})$.
The result is that $\mathcal{N}_{\vec{v}}$ \textit{inherits} the work applied to $\mathsf{parent}(\mathcal{N}_{\vec{v}})$, because an adversary attempting to remove the block from $\mathcal{N}_{\vec{v}}$ would need to have sufficient mining power to remove it from $\mathsf{parent}(\mathcal{N}_{\vec{v}})$.


\subsection{State Updates}
\label{sec:state}
In order to guarantee consistent application state between nodes, it is sufficient that all nodes in $\mathcal{N}_{\vec{v}}$ agree on the initial state (which can be defined in the genesis block) and then apply the same state updates to $\mathcal{S}_{\vec{v}}$ in the same order.
This is simple for transactions with the same origin and destination, as the state updates can be applied in the order that they are referenced by that blockchain. 
For transactions with different origin and destination, the state transition must be handled in two steps. 
First, $\mathcal{S}_{\vec{v_o}}$ is updated according to the transaction (e.g., the removal of an asset from the origin state) as soon as it is included in a block in the canonical chain of $\mathcal{B}_{\vec{v_o}}$. 
$\mathcal{S}_{\vec{v_d}}$, however, cannot be updated immediately, as there is initially no way for nodes operating $\mathcal{B}_{\vec{v_d}}$ to agree upon when the state transition should be applied. 
Protocol Rule \ref{rule:transition} describes the criteria which must be met for a state transition to be applied to $\mathcal{S}_{\vec{v_d}}$, and Protocol Rule \ref{rule:order} describes the order in which all state updates are applied when a block is processed.

\begin{definition}
\label{rule:transition}
Let $tx$ be a transaction with origin $\vec{v_o}$ and destination $\vec{v_d}$, and let $\vec{v_{a}}$ be the highest order common ancestor between $\vec{v_o}$ and $\vec{v_d}$.
The state transition pertaining to $\mathcal{S}_{\vec{v_o}}$ is applied as soon as the block containing $tx$ is a part of the canonical chain of $\mathcal{B}_{\vec{v_o}}$. 
The state transition pertaining to $\mathcal{S}_{\vec{v_d}}$, however, is only applied after a coincident block is found which is shared by $\mathcal{B}_{\vec{v_o}}$ and $\mathcal{B}_{\vec{v_{a}}}$ and, if $\mathsf{order}(\mathcal{B}_{\vec{v_{d}}}) > \mathsf{order}(\mathcal{B}_{\vec{v_{a}}})$, a subsequent coincident block is found which is shared by $\mathcal{B}_{\vec{v_d}}$ and $\mathcal{B}_{\vec{v_{a}}}$.
\end{definition}

Intuitively, a chain of coincident blocks must be constructed which link $\vec{v_o}$ and $\vec{v_d}$ through predecessor references, and if $\vec{v_o}$ and $\vec{v_d}$ are in different branches of the tree, the chain must travel ``up" and then ``back down" the hierarchy until that chain is established.
Nodes in $\mathcal{N}_{\vec{v_d}}$ can verify that the block containing $tx$ is in the canonical chain of $\mathcal{B}_{\vec{v_o}}$ because of the coincident block between $\mathcal{B}_{\vec{v_o}}$ and $\mathcal{B}_{\vec{v_{a}}}$, and they can also agree upon the existence of the coincident block between $\mathcal{B}_{\vec{v_d}}$ and $\mathcal{B}_{\vec{v_{a}}}$ and the updates to  $\mathcal{S}_{\vec{v_{d}}}$ which result from that coincident block. 

The next rule defines the order in which eligible state updates are to be applied, as all nodes must apply updates in the same order to guarantee consistent state.

\begin{definition}
\label{rule:order}
The state transitions for a given block $B$ in $\mathcal{B}_{\vec{v}}$ are applied to $\mathcal{S}_{\vec{v}}$ in the following order. 
First, if $B$ is a coincident block, all transactions with destination $\vec{v}$ which are eligible to be applied to $\mathcal{S}_{\vec{v}}$ according to Protocol Rule \ref{rule:transition} are applied in order by highest origin order to lowest origin order.
Transactions with the same origin order are applied in order of the blockchain index (i.e., from left to right in the tree) at that order, and transactions with the same origin index are applied in chronological order according to their inclusion in their origin blockchain. 
After that, all transactions directly referenced by $B$ are applied to $\mathcal{S}_{\vec{v}}$ in the order in which they are referenced by $B$.
\end{definition}

Protocol Rules \ref{rule:hlc}, \ref{rule:transition}, and \ref{rule:order} guarantee that all nodes in the same sub-network perform state transitions in the same order, meaning any two nodes which agree on the canonical chain will have consistent local state.
This is formalized by Theorem \ref{thm:ordering}.

\begin{theorem}
\label{thm:ordering}
For any given blockchain $\mathcal{B}_{\vec{v}}$, any two correct nodes in $\mathcal{N}_{\vec{v}}$ which agree on the canonical chain of $\mathcal{B}_{\vec{v}}$ will agree on $\mathcal{S}_{\vec{v}}$.
\end{theorem}


We prove Theorem \ref{thm:ordering} inductively, first remarking that any two correct nodes must agree upon the genesis block and the associated state when the blockchain is instantiated.
We then show the inductive step---for each block in the canonical chain of $\mathcal{B}_{\vec{v}}$, both nodes must apply the same state updates to $\mathcal{S}_{\vec{v}}$ and in the same order.
We prove this via contradiction, showing on a case by case basis that regardless of the origin and destination of a transaction, if one node applies the state update for that transaction and the other does not, then at least one of the nodes breaks a protocol rule. 
Therefore both nodes agree on the initial state and each subsequent state update, and the statement follows.
Due to space constraints, we omit the full proof from this paper.

As a corollary of Theorem \ref{thm:ordering}, all correct nodes in $\mathcal{N}_{\{1\}}$ which agree on the canonical chain of $\mathcal{B}_{\{1\}}$ also agree on $\mathcal{S}_{\{1\}}$. 
This ensures that all nodes in the system agree upon the state of the root blockchain in BlockReduce, thereby achieving similar guarantees to that of a single-blockchain system.


\section{Analysis}
\label{sec:perf}
In this section we show that BlockReduce achieves transaction throughput that scales superlinearly with the number of blockchains in each order.

\subsection{Decreased Propagation Delay via Network Partitioning}
\label{sec:decreased_delay}
Sub-networks of order $r > 1$ (i.e., all but the root sub-network) have a network delay $\Delta_r$ which is strictly less than $\Delta_1$, where $\Delta_1$ is analagous to the $\Delta$ experienced by a traditional blockchain operated by a full network of $N$ nodes. 
This is due to the decreased size of the sub-networks and the ability for nodes to select the sub-network with which they share low-latency peer connections in order to reduce the overall block propagation time that they experience. 

It is clear from Equation \ref{eq:delta_approx} that a smaller network naturally has lower propagation delays than a larger network. 
In order to enhance this intuition, if we assume that there are $q$ sub-networks of order $r$ and each sub-network is of equal size, then we can bound $\Delta_r$ as follows:

\begin{subequations}
\label{eq:subnetworks}
    \begin{align}
    \Delta_r &< \delta_r C_d \ln \big(\frac{N}{q}\big) \\
    &< \Delta_1 - \delta_r C_d \ln q
    \end{align}
\end{subequations}

Then the propagation delay for any order will be strictly increasing with the number of sub-networks in that order. 

Additionally, we remark that under the distribution of node-to-node latencies in Bitcoin as measured by \cite{Neudecker2019}, latency between nodes can vary significantly from one pair to the next. 
We assume that in the absence of a protocol mechanism requiring nodes to join a particular sub-network, each node will elect to mine in the sub-network which minimizes their peer-to-peer latencies to reduce the probability that their blocks are lost due to network forks.
As a result, we argue that in each sub-network of order greater than $1$, the average per-link propagation delay $\delta_r$ between nodes in order $r$ sub-networks will be smaller than the delay between nodes in $\mathcal{N}_{\{1\}}$---i.e.,  $\delta_r < \delta_1$ for all $r > 1$.
While we do not use this result in our proof, it nonetheless further supports the claim in Theorem \ref{thm:throughput}.

Overall, the network propagation delay within each sub-network will be much smaller than that of the network as a whole (i.e., the network delay of a similar system such as Bitcoin), and as a result the rate of block generation can be much higher within each sub-network. 

\subsection{Aggregate Throughput}
In this section, we show that within each order in the hierarchy, the total transaction throughput increases as the number of blockchains in that order increases even if the PoW efficiency remains fixed for each blockchain. 
Moreover, we show that the increase is superlinear with the number of blockchains in each order, as each additional blockchain further partitions the network and thus reduces the propagation delay experienced by nodes in each sub-network. 
We state this property of BlockReduce more formally in the following theorem.

\begin{theorem}
\label{thm:throughput}
The aggregate transaction throughput of each order in the BlockReduce protocol scales superlinearly with the number of blockchains in that order, and all blockchains in the hierarchy have identical PoW efficiency.
\end{theorem}

We prove this theorem by showing that the effective block generation rate of an order $r > 1$ blockchain grows as the number of order $r$ blockchains increases.

\begin{proof}

Recall that $\lambda_r$ is the total block generation rate of a blockchain of order $r$, then the effective block generation rate is $\lambda_r^* = \frac{\lambda_r}{1 + \lambda_r \Delta_r}$, and the PoW efficiency is $\mathcal{E}_r = \frac{1}{1 + \lambda_r \Delta_r}$. 
It suffices to show that if $q$ is the number of blockchains of order $r$, $\lambda_r^*$ increases as $q$ increases. 

We hold the PoW efficiency fixed between all blockchains to that of the root blockchain, i.e., $\frac{1}{1 + \lambda_1 \Delta_1} = \frac{1}{1 + \lambda_r \Delta_r}$
Substituting the results from Equation \ref{eq:subnetworks}, we get 

\begin{equation}
    \frac{1}{1 + \lambda_1 \Delta_1} > \frac{1}{1 + \lambda_r (\Delta_1 - \delta_r C_d \ln q)},
\end{equation}

which simplifies to 

\begin{equation}
    \lambda_r > \frac{\lambda_1 \Delta_1}{\Delta_1 - \delta_r C_d \ln q}.
\end{equation}

Then the effective block generation rate---i.e., the rate at which blocks are appended to the canonical chain---for an order $r$ blockchain with the same PoW efficiency as that of the root blockchain is

\begin{equation}
    \lambda_r^* > \frac{\lambda_1 \Delta_1}{(\Delta_1 - \delta_r C_d \ln q)(1 + \lambda_1 \Delta_1)}.
\end{equation}

Clearly, the right hand size of this equation increases with $q$, meaning $\lambda_r^*$ does as well. 
Then because all blocks are assumed to contain the same number of transactions, the statement follows. 
\end{proof}

\section{Discussion and Future Work}
\label{sec:conclusions}
\label{sec:conclusion}

BlockReduce is a PoW-based blockchain system which achieves high transaction throughput through a hierarchy of merged mined blockchains which each operate a partition of the overall application state in parallel. 
Critically, the full PoW available to the network is applied to all blockchains in BlockReduce, and cross-blockchain state transitions are enabled seamlessly within the core protocol.
In this section, we highlight several discussion points and avenues for future work.

\subsubsection{Self-selection of sub-network participation.} 
Mining nodes in BlockReduce are allowed to mine any vertical slice of blockchains within the PoW hierarchy tree. 
However, because the vast majority of miners in any PoW blockchain are economically motivated, most miners will elect to mine the blockchains which grant them the highest rewards. 
In this way, blockchains within each order will be self-balancing as miners naturally drift towards any available blockchains with reduced competition. 
In the absence of competitive advantage in mining power, miners will elect to mine the blockchains with which they share the lowest latency connections in order to minimize the probability that their blocks are lost to network forks. 
We believe that this alignment of incentives should result in the formation of low-latency clusters of mining nodes which are able to achieve very high PoW efficiency, although we leave the construction of a suitable incentive mechanism and a game-theoretical analysis to future work. 


\subsubsection{Low-value transaction security and settlement time tradeoffs.}
BlockReduce users have a high degree of flexibility when transacting, as they can control both the short-term security and the settlement time of their transactions by selecting which blockchain to transact in. 
Higher order blockchains have low settlement times and low security in the short term, as the sub-networks operating these blockchains have a high block generation rate but only control a fraction of the overall mining power. 
This is ideal for transactions of low economic value, as the consequence of a low-value transaction being removed from the canonical chain is minor. 
Users desiring a higher degree of security might elect to transact in lower order blockchains in order to utilize a larger fraction of the PoW of the network despite the tradeoff of longer settlement times. 
The precise characterization of this tradeoff will be specific to a particular implementation, and it is likely that any BlockReduce implementation will provide a default wallet which is capable of selecting the appropriate transaction location in an automated fashion.

\subsubsection{Expected hierarchy structure and future implementation.}
Although we have designed BlockReduce to support an arbitrary hierarchy tree, real-world hardware and network infrastructure as well as timing requirements of a functional blockchain network will limit both the number of orders and the number of blockchains per order that are realizable in practice. 
As the number of blockchains in the BlockReduce hierarchy increases, the time required for a cross-blockchain state transition also increases due to the decreasing relative frequency of coincident blocks linking each blockchain. 
For example, if a blockchain $\mathcal{B}$ has 3 child blockchains, then an average of 1 in every 3 blocks in $\mathcal{B}$ will be coincident with each child blockchain. 
If this number increases to 100 child blockchains, the expected time required for a coincident block to be found with one particular child increases dramatically.
Implementation and empirical testing will be required to determine the optimal configuration for any given use case and network topology.

\subsubsection{Conclusion}
We presented BlockReduce, a blockchain system which operates a hierarchy of blockchains in parallel, and showed that BlockReduce achieves transaction throughput that scales superlinearly with the number of blockchains operated by the system without reducing the security of each transaction.


\bibliographystyle{IEEEtran}
\bibliography{refs}
\vspace{12pt}

\end{document}